\documentstyle [twocolumn,epsf]{mn}
\newcommand{\mincir}{\raise
-3.truept\hbox{\rlap{\hbox{$\sim$}}\raise4.truept\hbox{$<$}\ }}
\newcommand{\magcir}{\raise
-3.truept\hbox{\rlap{\hbox{$\sim$}}\raise4.truept\hbox{$>$}\ }}
\newcommand{\minmag}{\raise
-3.truept\hbox{\rlap{\hbox{$<$}}\raise5.truept\hbox{$<$}\ }}
\newcommand{\be}{\begin{equation}}
\newcommand{\ee}{\end{equation}}
 \newcommand{\ba}{\begin{eqnarray}}
\newcommand{\ea}{\end{eqnarray}}
\newcommand{\brr}{\begin{array}}
\newcommand{\err}{\end{array}}
\newcommand{\bc}{\begin{center}}
\newcommand{\ec}{\end{center}}

\renewcommand{\baselinestretch}{1.0}

\title[Cosmological constrains from AGNs and SNIa data]
{Cosmological constrains from X-ray AGN clustering and SNIa data}
\author[Basilakos \& Plionis]{S. Basilakos$^{1}$ \&  M. Plionis$^{1,2}$.\\
\vspace{0.1cm}
$^1$ Institute of Astronomy \& Astrophysics, National Observatory of Athens, 
I. Metaxa \& V. Pavlou, Palaia Penteli, 15236 Athens, Greece \\
$^2$ Instituto Nacional de Astrofisica, Optica y Electronica (INAOE)
Apartado Postal 51 y 216, 72000, Puebla, Pue., Mexico
}

\begin{document}

\maketitle

\begin{abstract}
We put constraints on the main cosmological parameters of 
different spatially flat cosmological 
models by combining the recent clustering results of XMM-{\it
Newton} soft (0.5-2\,keV) X-ray sources, which have a redshift
distribution with median redshift $z\sim 1.2$, and SNIa data.
Using a likelihood procedure we find that the model which
best reproduces the observational data and that is consistent with stellar ages
is the {\em concordance} $\Lambda$CDM model with: 
$\Omega_{\rm m}\simeq 0.28$, w$\simeq -1$, $H_{\circ}\simeq 72$ km
s$^{-1}$ Mpc$^{-1}$, $t_{\circ} \simeq 13.5$ Gyr and has an X-ray 
AGN clustering evolution which is constant in physical coordinates.
For a different clustering evolution model (constant in comoving coordinates)
we find another viable model, although less probable due to the
smaller age of the Universe, with
$\Omega_{\rm m}\simeq 0.38$, w$\simeq -1.25$,
$H_{\circ}\simeq 70$ km s$^{-1}$ Mpc$^{-1}$ and $t_{\circ} \simeq 12.9$ Gyr. 
\end{abstract}


\vspace{1.0cm}

\section{Introduction}
Recent advances in observational cosmology, based on the analysis of
a multitude of high quality observational data (Type Ia
supernovae, cosmic microwave background (CMB), 
large scale structure, age of the
globular clusters, high redshift galaxies),
strongly indicated that we are living in a flat
($\Omega_{\rm tot}=1)$ accelerating Universe 
containing a small baryonic component, non-baryonic
cold dark matter (CDM) to explain the clustering of
extragalactic sources and an extra component with
negative pressure, usually named ``dark energy",
to explain the present accelerated expansion of the universe
(eg. Riess, et al. 1998; Perlmutter et al.
1999; Efstathiou et al. 2002; Spergel et al. 2003; Percival et
al. 2002; Tonry et al. 2003; Schuecker et al. 2003; Riess et al. 2004; 
Tegmark et al. 2004)

The last few years there have been many theoretical 
speculations regarding the nature 
of the exotic ``dark energy''. Various candidates 
have been proposed in the literature, among which 
the time varying $\Lambda$-parameter (eg. Ozer \& Taha 1987), 
a scalar field having a self-interaction potential $V(\Phi)$ 
with the field energy density decreasing with a slower rate than the
matter energy density (dubbed also ``quintessence'', 
eg. Peebles \& Ratra 2003 and references therein) or an 
extra ``matter'' component, which is described by an equation of state
$p_{Q}={\rm w}\rho_{Q}$ with w$<-1/3$ (a redshift dependence of w is
also possible but present measurments are not precise enough to allow
meaningful constraints; eg. Dicus \& Repko 2004). A particular case of
``dark energy'' is the traditional 
$\Lambda$-model which corresponds to w$=-1$.
Note that a variety of observations indicate that 
w$< -0.6$ for a flat geometry (eg. Ettori, Tozzi \& Rosati 2002; 
Tonry et al. 2003; Riess et al. 2004; Schuecker 2005).

In this paper we put constraints on spatially flat cosmological models
using the recently derived clustering properties 
of the XMM-{\it Newton} soft (0.5-2\,keV)
X-ray point sources (Basilakos et al. 2005), 
the SNIa data (Tonry et al. 2003) and the age of
globular clusters (eg. Krauss 2003; Cayrel et al. 2001).
Hereafter will use the normalized Hubble 
constant $H_{\circ}=100 \;h$ km s$^{-1}$ Mpc$^{-1}$.

\section{X-ray AGN Clustering}
In a previous paper (Basilakos et al 2005) we derived the 
angular correlation function of the soft (0.5-2\,keV) XMM 
X-ray sources using a shallow (2-10\,ksec) wide-field survey ($\sim
2.3$ deg.$^{2}$).
A full description of the data reduction, source detection and flux 
estimation are presented in Georgakakis et al. (2004). Note that the 
survey contains 432 point sources within an effective
area of $\sim 2.1$ deg$^{2}$ (for $f_x \ge 2.7 \times
10^{-14}$ erg cm$^{-2}$ s$^{-1}$ ), while for 
$f_x \ge 8.8 \times 10^{-15}$ erg cm$^{-2}$ s$^{-1}$ the effective
area of the survey is $\sim 1.8$ deg$^{2}$.
In Basilakos et al (2005)
we present the details of the correlation function estimation, the
various biases that should be taken into account (the amplification
bias and integral constraint), the survey luminosity
and selection functions as well as issues related to possible stellar
contamination. 
In particular, the redshift selection function of our X-ray sources, 
derived by using the soft-band luminosity function of Miyaji, Hasinger \& Schmidt
(2000), and assuming the realistic luminosity dependent density 
evolution of our sources, predicts a characteristic depth of $z\simeq 1.2$ 
for our sample (for details see Basilakos et al. 2005).

Our aim here is to compare the theoretical clustering predictions from
different flat cosmological models to the actual observed angular
clustering of distant X-ray AGNs.
For the purpose of this study we use Limber's formula
which relates the angular, $w(\theta)$, and the spatial, $\xi(r)$,
correlation functions. In the case of a spatially flat Universe,
Limber's equation can be written as:
\be 
w(\theta)=2\frac{\int_{0}^{\infty} \int_{0}^{\infty} x^{4} 
\phi^{2}(x) \xi(r,z) {\rm d}x {\rm d}u}
{[\int_{0}^{\infty} x^{2} \phi(x){\rm d}x]^{2}} \;\; , 
\ee
where $\phi(x)$ is the selection function (the probability 
that a source at a distance $x$ is detected in the survey) and 
$x$ is the coordinate distance related to the redshift through 
\be
x(z)=\frac{c}{H_{\circ}} \int_{0}^{z} \frac{{\rm d}y}{E(y)}\;\; ,
\ee
with $E(z)=[\Omega_{\rm m}(1+z)^{3}+(1-\Omega_{\rm m})(1+z)^{\beta}]^{1/2}$
and $\beta=3(1+{\rm w})$. The number of objects within 
a shell $(z,z+{\rm d}z)$ and in a given 
survey of solid angle $\omega_{s}$ is: 
\be
\frac{{\rm d}N}{{\rm d}z}=\omega_{s}
x^{2} n_s \phi(x)\left(\frac{c}{H_{\circ}}\right) E^{-1}(z)\;\;.
\ee
where $n_s$ is the comoving number density at zero redshift.
Combining the above system of  equations we obtain:
\begin{equation}\label{eq:angu}
w(\theta)=2\frac{H_{\circ}}{c} \int_{0}^{\infty} 
\left(\frac{1}{N}\frac{{\rm d}N}{{\rm d}z} \right)^{2}E(z){\rm d}z 
\int_{0}^{\infty} \xi(r,z) {\rm d}u
\end{equation} 
where $r$ is the physical separation between two sources,  
having an angular separation, $\theta$, given by $r \simeq 
 (1+z)^{-1} \left(u^{2}+x^{2}\theta^{2} \right)^{1/2}$
(small angle approximation).
Therefore, in order to estimate the expected $w(\theta)$ in a
cosmological model we also need to determine 
the source redshift distribution ${\rm d}N/{\rm d}z$, which as we said
previously, is estimated by integrating the appropriate Miyaji et al. (2000)
luminosity function.

\subsection{The Role and Evolution of Galaxy Bias}
It has been claimed that the large scale 
clustering of different mass tracers (galaxies or clusters) 
is biased with respect to the matter distribution (cf. Kaiser 1984; 
Bardeen 1986). It is also 
an essential ingredient for cold dark matter (CDM) 
models to reproduce the observed galaxy 
distribution (cf. Davis et al. 1985).
Within the framework of
linear biasing (cf. Kaiser 1984; Benson et al. 2000), the mass-tracer 
and dark-matter spatial correlations, at some redshift $z$, are related by:
\be
\label{eq:spat}
\xi(r,z)=\xi_{\rm DM}(r,z)b^{2}(z) \;\;, 
\ee
where $b(z)$ is the bias evolution function.
This has been shown to be 
a monotonically increasing function of redshift
(Mo \& White 1996; Matarrese et al. 1997; Basilakos \& Plionis 2001
and references therein). Here we use the bias evolution model of 
Basilakos \& Plionis (2001, 2003), 
which is based on linear perturbation theory and the 
Friedmann-Lemaitre solutions of the cosmological
field equations. We remind the reader that
for the case of a spatially flat cosmological model
our general bias evolution can be written as:
\be\label{eq:88} 
b(z)= {\cal A} E(z)+{\cal C} E(z) \int_{z}^{\infty}
\frac{(1+y)^{3}}{E^{3}(y)} {\rm d}y +1 \;\;.
\ee
Note that our model gives a family of bias
curves, due to the fact that it has two unknowns
(the integration constants ${\cal A},{\cal C}$). 
In this paper, for simplicity, we 
fix the value of ${\cal C}$ to $\simeq 0.004$, as was determined
in Basilakos \& Plionis (2003) from the {\it 2dF} galaxy correlation
function. We have tested the robustness of our results
by increasing ${\cal C}$  by a factor of 10 and 100 to find 
differences of only $\sim$ 5\% in the fitted 
values of $\Omega_{m}$ and $b_{\circ}$.
This behaviour can be explained from the fact that the dominant term
in the right hand side of eq. (6) is the first term [$\propto
(1+z)^{3/2}$] while the second term has a slower dependence on redshift 
[$\propto (1+z)$].

\subsection{Clustering Evolution}
The redshift evolution of the spatial mass correlation function,
$\xi_{\rm DM}(r,z)$, can be written as the Fourier transform of the 
spatial power spectrum $P(k)$. Using also eq. (5) we have:
\be
\label{eq:spat1}
\xi(r,z)=\frac{(1+z)^{-(3+\epsilon)}b^{2}(z)}{2\pi^{2}}\int_{0}^{\infty} k^{2}P(k) 
\frac{{\rm sin}(kr)}{kr}{\rm d}k \;\;,
\ee
where $k$ is the comoving wavenumber. 
Note that the parameter $\epsilon$ parametrizes the type 
of clustering evolution (eg. de Zotti et al. 1990). 
If $\epsilon=\gamma-3$ (with $\gamma$ the slope of the 
spatial correlation function; $\gamma=1.8$)
the clustering is constant in comoving 
coordinates while if $\epsilon=-3$ the 
clustering is constant in physical coordinates.

The power spectrum of our CDM models is given by $P(k) \propto k^{n} T^{2}(k)$
with scale-invariant ($n=1$) primeval 
inflationary fluctuations and $T(k)$ the CDM transfer function. 
In particular, we use the transfer function parameterization as in
Bardeen et al. (1986), with the corrections given approximately
by Sugiyama (1995) while the
normalization of the power spectrum is given by: $\sigma_8 \simeq 
0.5 \Omega_{\rm m}^{-\gamma}$ with $\gamma \simeq 0.21-0.22 {\rm w} +0.33
\Omega_{\rm m}$ (Wang \& Steinhardt 1998). We caution that
this fit, based on the rich cluster abundances, has been derived for
w$\ge -1$. In this work we assume that the fit is valid also for
w$<-1$.
Note that we also use the
non-linear corrections introduced by Peacock \& Dodds (1994).  

\section{Cosmological Constraints}

\subsection{X-ray AGN Clustering likelihood}
It has been shown that the
application of the correlation function analysis on samples of
high redshift galaxies can be used as a useful tool 
for cosmological studies (eg. Matsubara 2004).
In our case to constrain the cosmological parameters
we utilize a standard $\chi^{2}$ 
likelihood procedure to compare the measured 
XMM soft source angular correlation function (Basilakos et al. 2005) 
with the prediction of different spatially flat cosmological models.
In particular, we define the likelihood estimator\footnote{Likelihoods
  are normalized to their maximum values.} as:
${\cal L}^{\rm AGN}({\bf c})\propto {\rm exp}[-\chi^{2}_{\rm AGN}({\bf c})/2]$
with:
\be
\chi^{2}_{\rm AGN}({\bf c})=\sum_{i=1}^{n} \left[ \frac{ w_{\rm th}
(\theta_{i},{\bf c})-w_{\rm obs}(\theta_{i}) }
{\sigma_{i}} \right]^{2} \;\;.
\ee 
where ${\bf c}$ is a vector containing the cosmological 
parameters that we want to fit and $\sigma_{i}$ the observed angular 
correlation function uncertainty. 
Here we work within the framework 
of a flat ($\Omega_{\rm tot}=1$) cosmology 
with primordial adiabatic fluctuations and baryonic
density of $\Omega_{\rm b} h^{2}\simeq 0.022$ 
(eg. Kirkman et al. 2003). In this case the corresponding vector
is ${\bf c}\equiv (\Omega_{\rm m},{\rm w},h,b_{\circ})$.   
We sample the various parameters as follows:
the matter density $\Omega_{\rm m} \in [0.01,1]$ in steps of
0.01; the equation of state parameter w$\in [-3,-0.35]$ in steps
of 0.05, the dimensionless Hubble constant $h \in [0.5,0.9]$ in steps
of 0.02 and the X-ray sources bias at the present time
$b_{\circ} \in [0.5,4]$ in steps of 0.05.
Note that in order to investigate possible equations of state,
we have allowed the parameter w to take values below -1. Such models 
correspond to the so called {\em phantom} cosmologies (eg. Caldwell 2002;
Corasaniti et al. 2004).

The resulting best fit parameters, for the two
clustering evolution models, are presented in the first two rows of
Table 1.  
It is important to note that our estimates for the Hubble parameter $h$
are in very good agreement with those derived ($h=0.72\pm
0.07)$ by the HST key project (Freeman et al. 2001).
In Fig.1 we present the 1$\sigma$, 2$\sigma$ and $3\sigma$
confidence levels in the $({\rm w},h)$ and $({\rm w},b_{\circ})$ planes by 
marginalizing the former over $\Omega_{\rm m}$ and $b_{\circ}$ and the
latter over $\Omega_{\rm m}$ and $h$. It is evident that w is
degenerate with respect to both $h$ and the bias at the present time.
\begin{figure}
\mbox{\epsfxsize=8.3cm \epsffile{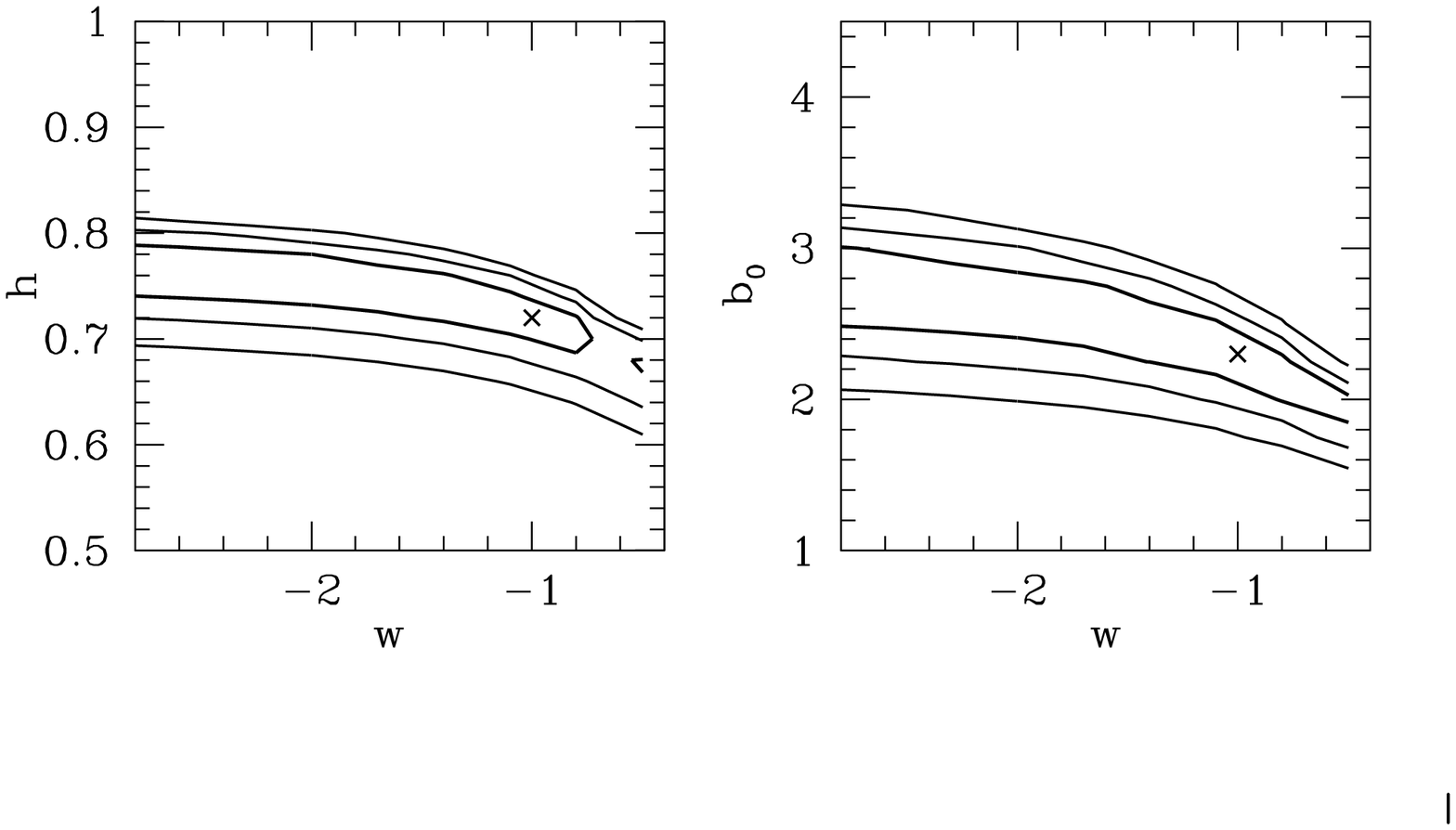}}
\caption{Likelihood contours in the $({\rm w},h)$ plane (left panel)
and the $({\rm w},b_{\circ})$ plane (right panel) for
$\epsilon=-1.2$ (a similar degeneracy is true also for the 
$\epsilon=-3$ clustering evolution model). The 
contours are 
plotted where $-2{\rm ln}{\cal L}/{\cal L}_{\rm max}$ is equal
to 2.30, 6.16 and 11.83, corresponding 
to 1$\sigma$, 2$\sigma$ and 3$\sigma$ confidence level.}
\end{figure}

Figure 2 also shows 
the 1$\sigma$, 2$\sigma$ and 3$\sigma$
confidence levels (continuous lines) in the $(\Omega_{\rm m},{\rm w})$ plane by 
marginalizing over the Hubble constant
and the bias factor at the present time
\footnote{Hereafter, when we marginalize over the 
Hubble constant we will use $h=0.72$ for
$\epsilon=-1.2$ and $h=0.7$ for $\epsilon=-3$.}. 
The equation of state parameter likelihood is not 
constrained by this analysis and
all the values in the interval 
$-3\le {\rm w} \le -0.35$ are acceptable
within the $1\sigma$ uncertainty. 
Therefore, in order to put further constraints on 
w we additionally use the SNIa data (Tonry et al. 2003)
as well as the so called age limit, given by the age of the oldest
globular clusters in our Galaxy ($t_{\circ}> 12.7$ Gyr;  Krauss 2003; 
Cayrel et al. 2001 and references therein).

\begin{table*}
{\small 
\caption[]{Cosmological parameters from the likelihood analysis:
The 1$^{st}$ column indicates the data used (the last two rows correspond to
the joint likelihood analysis). Errors of the fitted parameters 
represent $1\sigma$ uncertainties. Note that for the joined
analysis the corresponding results are marginalized over 
the Hubble constant and the bias factor at the present time, for which
we use the values indicated.}
\tabcolsep 10pt
\begin{tabular}{ccccccc} 
\hline
Data& $\Omega_{\rm m}$& w& $h$& $b_{\circ}$& $t_{\circ}$& $\chi^{2}/{\rm dof}$ \\ \hline 
{\rm XMM} $(\epsilon=-1.2$) &  $0.31^{+0.16}_{-0.08}$  & uncons. (w$=-1$) & $0.72^{+0.02}_{-0.18}$&  $2.30^{+0.70}_{-0.20}$ &13.0&0.82\\
{\rm XMM}$(\epsilon=-3.0$) & $0.38^{+0.02}_{-0.14}$    & uncons. (w$=-1$) &$0.70^{+0.04}_{-0.16}$&  $1.20^{+0.60}_{-0.30}$&12.6&0.84\\
{\rm XMM}$(\epsilon=-1.2$)/{\rm SNIa}& $0.28\pm 0.02$  & $-1.05^{+0.10}_{-0.20}$& $0.72$& $2.30$&13.5& 0.87\\
{\rm XMM}$(\epsilon=-3.0$)/{\rm SNIa}& $0.38\pm {0.03}$& $-1.25^{+0.10}_{-0.25}$& $0.70$&  $1.20$&12.9& 0.85\\ \hline
\end{tabular}
}
\end{table*}
 
\subsection{The likelihood from the SNIa}
We use the sample of 172 supernovae of Tonry et al. (2003) in order
to constrain $\Omega_{\rm m}$ and the
equation of state in the framework of a flat geometry ($\Omega_{\rm tot}=1$).
In this case, the corresponding vector ${\bf c}$
is: ${\bf c}\equiv (\Omega_{\rm m},{\rm w})$ and
the likelihood function can be written as: 
${\cal L}^{\rm SNIa}({\bf c})\propto {\rm exp}[-\chi^{2}_{\rm SNIa}({\bf c})/2]$
with:
\be
\chi^{2}_{\rm SNIa}({\bf c})=\sum_{i=1}^{172} \left[ \frac{ {\rm log}
    D^{\rm th}_{\rm L}
(z_{i},{\bf c})-{\rm log}D^{\rm obs}_{\rm L}(z_{i}) }
{\sigma_{i}} \right]^{2} \;\;.
\ee 
where $D_{\rm L}(z)$ is the dimensionless luminosity
distance, $D_{\rm L}(z)=H_{\circ}(1+z)x(z)$
and $z_{i}$ is the observed redshift. 
The green lines in Fig. 2 represents 
the $1\sigma$, $2\sigma$,and $3\sigma$,  
confidence levels in the $(\Omega_{\rm m},{\rm w})$ plane.
We find that the best fit solution is $\Omega_{\rm m}=0.30 \pm 0.04$ 
for w$>-1$, in complete agreement with 
previous SNIa studies (Tonry et al. 2003; Riess et al. 2004).

\subsection{The joined likelihoods}
In order to combine the X-ray AGN clustering properties 
with the SNIa data we perform a joined likelihood analysis and
marginalizing the X-ray clustering results 
over $h$ and $b_{\circ}$, which are not constrained by the value of w
(see Fig.1), we obtain:
${\cal L}^{\rm joint}(\Omega_{\rm m},{\rm w})={\cal L}^{\rm AGN} \times 
{\cal L}^{\rm SNIa}$.
Taking also into account the age limit (the yellow area in Fig. 2)
the likelihood for the $\epsilon=-1.2$ clustering evolution model peaks at 
$\Omega_{\rm m}=0.28 \pm 0.02$ with w$=-1.05^{+0.1}_{-0.2}$
(corresponding to $t_{\circ}=13.5$ Gyr) which
is in excellent agreement with the WMAP results of Spergel et al. (2003)
and the REFLEX X-ray clusters + SNIa results of Schuecker et
al. (2003). For the  $\epsilon=-3$ clustering evolution model we obtain
$\Omega_{\rm m}=0.38 \pm 0.03$ with w$=-1.25^{+0.10}_{-0.25}$
(which corresponds to $t_{\circ}=12.9$ Gyr).
The latter model appears to be
marginally ruled out by the stellar ages. Note that the normalization
of the power spectrum that corresponds to these models is
$\sigma_{8}\simeq 0.98$ and 0.90, respectively.

\begin{figure}
\mbox{\epsfxsize=8.5cm \epsffile{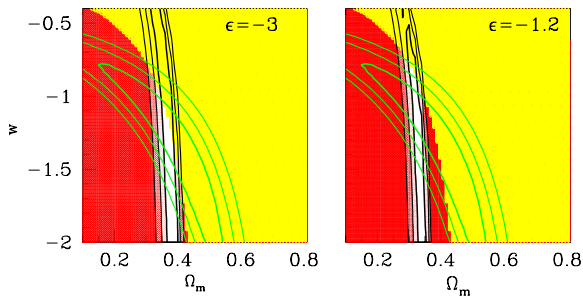}}
\caption{Likelihood contours in the $(\Omega_{\rm m},{\rm w})$ plane. The 
contours correspond to 1$\sigma$ (2.30), 2$\sigma$ (6.16) and 3$\sigma$
(11.83) confidence levels,
using the two different clustering 
behaviors (left panel for $\epsilon=-3$ and right panel for 
$\epsilon=-1.2$).
Note that the black and the green 
lines correspond to the X-ray AGN clustering and SNIa results, 
respectively while the yellow area is ruled out by the stellar ages.}
\end{figure}

It is evident that the combined likelihood analysis puts strong 
constraints on the value of w and once including stellar ages it
appears to favor the standard {\em concordance} 
$\Lambda$CDM ($\Omega_{\rm} \simeq 0.3$, w$\simeq -1$) cosmological model as well as a comoving 
AGN clustering model ($\epsilon=-1.2$).
However, the model with w$\simeq -1.25$, $\Omega_{\rm m} \simeq
0.38$ and $\epsilon=-3$ cannot be ruled out at any significant level.

Many other recent analyses utilizing different combinations of data
seem to agree with the former cosmological model.
For example, Tegmark et
al. (2004) used the WMAP CMB anisotropies 
in combination with the SDSS galaxy power spectrum and found a good
 $\Lambda$CDM fit with $\Omega_{\rm m}=0.30 \pm 0.04$ and
$h=0.70^{+0.04}_{-0.03}$ (see also 
Spergel et al. 2003; Percival et al. 2003; Schuecker et al. 2003).
Also combining the gas fraction in relaxed X-ray luminous 
clusters with the CMB and SNIa has provided stringent
constrains with $\Omega_{\rm m}\simeq 0.3$ and $w\simeq -1$
(eg. Allen et al. 2004; Rapetti, Allen \& Weller 2004). 

\section{Conclusions}
We have combined for the first time the clustering properties
of distant X-ray AGNs, identified as soft (0.5-2 keV) point sources in
a shallow $\sim$ 2.3 deg$^{2}$ XMM survey, which have a $z$-distribution
that peaks at $z\simeq 1.2$,
 with the SNIa data. From the X-ray AGN clustering likelihood analysis alone
we constrain $h\simeq 0.72 \pm 0.03$ (where the uncertainty is found
after marginalizing over w and $b_{\circ}$). From the 
joined likelihood analysis and taking into account stellar ages
we constrain the matter density and the equation of state parameters.
The best model appears to be one with
$\Omega_{\rm m}\simeq 0.28$, w$\simeq -1$ and a
stable in comoving coordinates X-ray 
AGN clustering model. However, the model with
$\Omega_{\rm m}\simeq 0.38$, w$\simeq -1.25$ and
constant in physical coordinates ($\epsilon=-3$) X-ray 
AGN clustering, of which the predicted age is
marginally consistent with stellar ages, 
cannot be excluded at any significant level by our analysis.

\end{document}